\documentclass[aps,amssymb,amsmath,prl,twocolumn,superscriptaddress]{revtex4-1}

\usepackage{graphicx}
\usepackage{epstopdf}
\usepackage[all]{xy}

\usepackage{amssymb}
\usepackage{epstopdf}
\usepackage{amsmath}
\usepackage{amsthm}
\usepackage[extension=xxx]{hyperref}
\newcommand{\unit}[1]{\ensuremath{\, \mathrm{#1}}}
\usepackage{color}

\def\be{\begin{equation}}   \def\ee{\end{equation}}
\def\eq#1{{Eq.\ref{#1}}}    \def\fig#1{{Fig.\ref{#1}}}

\def\kT{k_{{}_{\rm B}}T}

\begin{document}

\title{Transient domain formation in membrane-bound organelles undergoing maturation}
\author{Serge Dmitrieff, Pierre Sens\\ \vspace{0.2cm}{\em {\small Laboratoire Gulliver (CNRS UMR 7083)\\ ESPCI, 10 rue Vauquelin, 75231 Paris Cedex 05 - France\\ serge.dmitrieff@espci.fr, pierre.sens@espci.fr}}}
\date{August 25, 2013}

\begin{abstract}
The membrane components of cellular organelles have been shown to segregate into domains as the result of biochemical maturation. We propose that the dynamical competition between maturation and lateral segregation of membrane components regulates domain formation. We study a two-component fluid membrane in which enzymatic reaction irreversibly converts one component into another, and  phase separation triggers the formation of transient membrane domains. The maximum  domains size is shown to depend on the maturation rate as a power-law similar to the one observed for domain growth with time in the absence of maturation, despite this time dependence not being verified in the case of irreversible maturation. This control of domain size by enzymatic activity could play a critical role in intra-organelle dynamics.
\end{abstract}

\maketitle

\section{Introduction}

Molecules secreted and internalized by Eukaryotic cells follow well defined routes, the secretory or endocytic pathways, along which they are exposed to a succession of biochemical environments by sequentially visiting different membrane-bound organelles \cite{kelly1985pathways}. Different organelles have different membrane compositions, as well as a distinct set of membrane-associated proteins, refered to henceforth as the membrane {\em identity}. Interestingly, it has been shown that the identity of some organelles changes with time ; for example, the early endosome (a compartment digesting newly internalized content) has a different identity from the late endosome, which then becomes a lysosome \cite{zerial:2001}. One fundamental issue underlying the organization of  intracellular transport  is whether progression along the various pathways occurs by exchange between organelles of fixed  biochemical identities (via the budding and scission of carrier vesicles), or by the biochemical maturation of the organelles themselves \cite{kelly1985pathways,zerial:2001}. This question is particularly debated for the Golgi apparatus, where proteins undergo post-transcriptional maturation and sorting. The Golgi is divided into early ({\em cis}), middle ({\em medial}) and late ({\em trans})  $\unit{\mu m}$-size compartments called cisternae.  In yeast, each  cisterna appears to undergo independent biochemical maturation from a {\em cis} to a {\em trans} identity in less than 1 min \cite{matsuura:2006,losev:2006}. In higher eukaryotes, the cisternae form a tight and polarized stack with {\em cis} and {\em trans} ends, through which proteins travel in  about 20 min \cite{emr:2009}. Whether transport through the stack occurs by inter-cisternal exchange  or by the maturation of entire cisternae remains controversial \cite{emr:2009}.

Maturation in an organelle membrane causes  different membrane identities to transiently coexist and may trigger the formation of transient membrane domains. Membrane components have indeed been seen to segregate into domains in both yeast \cite{matsuura:2006,losev:2006} and  mammalian Golgi cisternae \cite{patterson:2008}.
 This is the case of proteins of the Rab family, though to be essential identity labels of cellular organelles \cite{zerial:2001}. The so-called {\em Rab cascade}, in which the activation of one Rab inactivates the preceding Rab along the pathway, is thought to permit the  sequential maturation of the organelle identity \cite{rivera:2009}. Domains could also emerge from the maturation of ceramids (present in cis-Golgi) into sphingomyelin (present in trans-Golgi), as these two species are known to lead to domain formation on vesicles \cite{sot:2006}. Finally, there is a continuous gradient of membrane thickness from cis to trans Golgi compartments \cite{mitra:2004} and thickness mismatch can lead to phase separation in model membranes \cite{heberle:2013}.
 
  It has been argued that membrane domains in organelles could undergo budding and scission, and hence control inter-organelle transport \cite{pfeffer:2010b}. This raises the interesting possibility that the rate of domain formation could control the rate of transport. To quantitatively assess this possibility, we studied transient domain formation in an ideal two-component membrane. We consider an irreversible transformation (maturation) $A\rightarrow B$ taking place between two components, with $A$ and $B$ representing distinct biochemical identities, and we investigate the phase behavior of a such membrane.

The kinetics of phase separation in binary mixtures have been abundantly studied \cite{bray:1994}. In the context of fluid membranes, hydrodynamic flows in the membrane and the surrounding media make the problem quite complex. Several dynamical regimes have been reported, and a unified picture of has not yet emerged  \cite{camley:2010,fan:2010}. For deformable fluid membranes such as cellular membranes, the budding of membrane domains \cite{lipowsky:1992} makes the dynamics of phase separation even more complex \cite{sunil:1998,sunil:2001,laradji:2004}. Here, we study transient phase separation on flat membranes, and we implement membrane deformability at a phenomenological level by introducing a critical domain size beyond which flat domains are unstable. If domains reach such a size, they undergo a budding transition and may serve as transport intermediates, provided a scission mechanism ({\em e.g} the activity of specialized proteins such as dynamin \cite{ferguson:2012}) separates budded domains from the rest of the membrane.

The budding of membrane domains may for instance be driven by the line energy associated with domain boundaries \cite{lipowsky:1992}, expressed as the domain line tension $\gamma$ times the boundary length. Budding is resisted by the membrane bending rigidity $\kappa$ and surface tension $\sigma$, and will occur for a finite range of domain size $R$ \cite{pio_regul}:\begin{equation}
4 \frac{\kappa}{\gamma} < R < 2 \frac{\gamma}{\sigma}\, .
\label{range}
\end{equation}
For typical values of the parameters: $\kappa\simeq10\kT$ and $\gamma\simeq1\unit{pN}$, the lower bound is a fraction of the typical  size of Golgi cisternae ($\sim\unit{\mu m}$), and the scenario of a line tension induced domain pinching appears realistic. The upper bound could be restrictive in a system with low area/volume ratio, where pinching might increase membrane tension and lead to incomplete buds. This is often observed in artificial vesicles, but this constraint does not appear stringent in organelles such as the Golgi. We assume henceforth that the rate of membrane deformation is much faster than the rate of chemical maturation, so that only the lower bound of \eq{range} is relevant.

In this article, we show that, in a membrane undergoing irreversible maturation, the maximal size of transient domains follows a power law with respect to the maturation rate. First, we illustrate two modes of domain growth in fluid membranes. We then show the influence of maturation on domain growth, and the predicted power laws are then confirmed numerically. Because budding depends on domain size, this means that organelle transport can be controlled by the maturation rate.

\section{Phase separation kinetics in fluid membranes}

 In order to elucidate the role of maturation, we only consider phase separation in a flat fluid membrane and we disregard the influence of the surrounding fluid. This approximation is valid for domains smaller than $\sim\eta_2/\eta_3$, where $\eta_3$ is the three-dimensional viscosity of the surrounding fluid (the cytoplasm and the lumen of cellular organelles in the cell), and $\eta_2$ is the two-dimensional viscosity of the membrane \cite{saffman:1975}. For biological membranes, one expects $\eta_2/\eta_3\gtrsim\unit{\mu m}$.

Domain growth is initially dominated by the so-called ``Ostwald ripening'', where large domains grow by adsorbing diffusing matter evaporated from smaller domains. At later time, hydrodynamic effects, leading in particular to domain coalescence, dominate the growth. In both cases, after an early nucleation stage, the domain size $R_c$ increases with time according to a power law \cite{bray:1994}:
\begin{eqnarray}
R_c^{1/\alpha-1} \dot{R_c} \propto \tilde D(\bar\phi) \quad:\quad R_c\sim t^\alpha.
\label{powerlaw}
\end{eqnarray}
The  exponent $\alpha$ and the transport coefficient $\tilde D$ depend on the dominant growth process, and the latter also depends on composition, line tension, and component mobility. 

\subsection{Thermodynamics}

The thermodynamics of phase separation in a two-component membrane containing distinct biochemical identities $A$ and $B$ can be studied using a local order parameter $\phi$ varying continuously between  $\phi=0$ for A-rich and $\phi=1$ for B-rich membrane regions. We use the classical Landau free energy \cite{bray:1994}:
\begin{eqnarray}
&\mathcal{F}=\int d^2 \mathbf{r} \left[ V [\phi(\mathbf{r})] + \frac{1}{2} \zeta\| \boldsymbol{\nabla} \phi \|^2 \right] \label{compofreeE} \\
&V [\phi] =\frac{\kT}{a^2}\left(\phi \log{\phi} + (1-\phi) \log{(1-\phi)}\right)+ \frac{K}{2a^2} \phi ( 1- \phi), \nonumber
\end{eqnarray}
where $a$ is a molecular size. This energy is the sum of an interfacial term (of parameter $\zeta$) and a potential term $V$ that includes the translational entropy and the interaction between the two phases (repulsive if $K>0$).

Phase separation occurs spontaneously inside the spinodal region of the phase diagram, defined by $K> \kT/(\bar{\phi} (1-\bar{\phi}))$, where $\bar{\phi}$ is the mean  value of $\phi$ in the system \cite{chaikin:1995}. Within this region, the interface between A-rich and B-rich domains is sharp and the energy of interaction reduces to a line energy characterised by the line tension $\gamma$ (see Supplementary Information - S.I. - for more details). 

\begin{figure}[b] 
   \includegraphics[width=7.5cm]{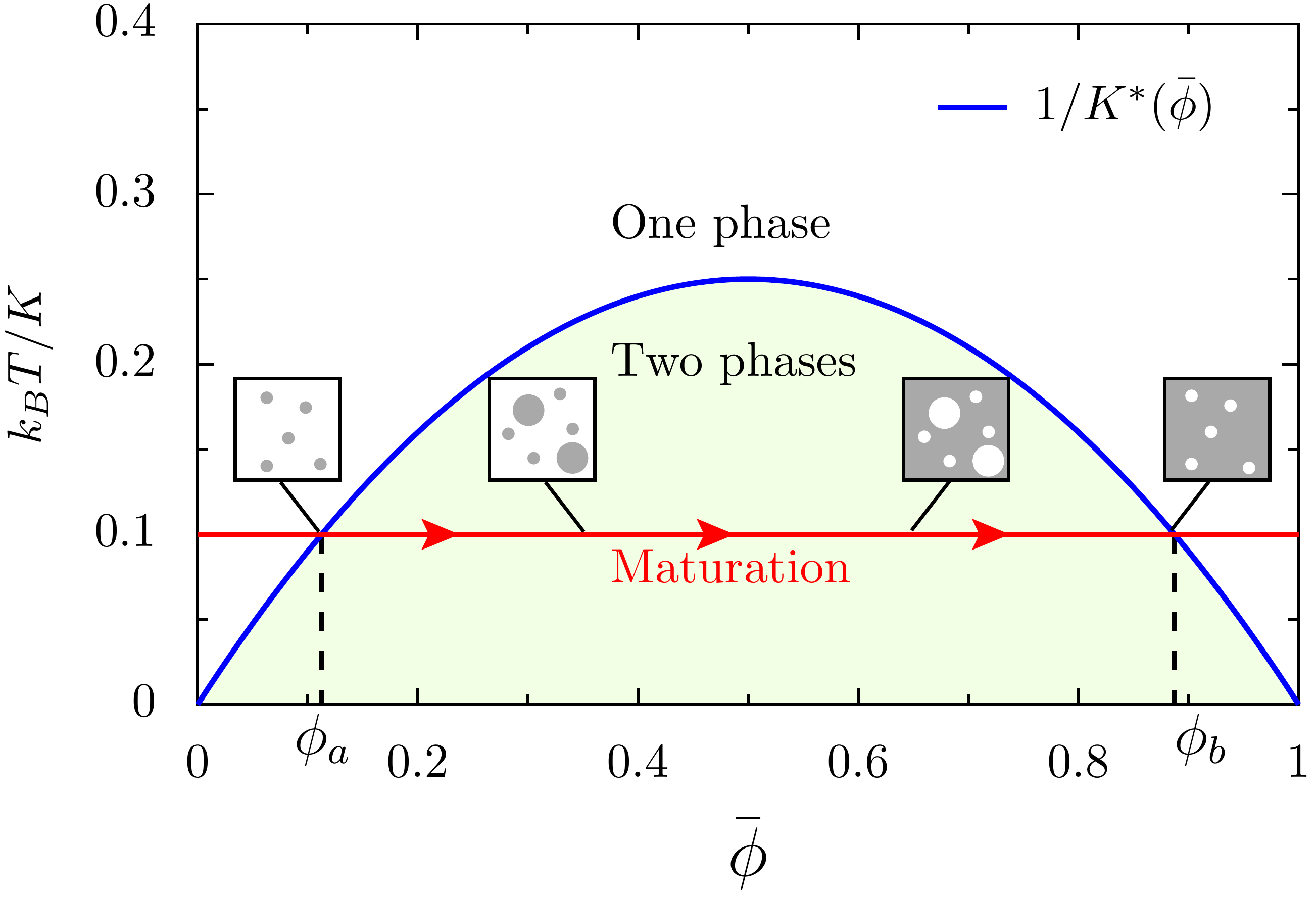} 
   \caption{\small (color online) Phase diagram of a binary mixture undergoing chemical maturation.  The blue line represents the spinodal line. The chemical reaction (maturation) increases the mean order parameter $\bar{\phi}$ from $0$ to $1$.}
   \label{spinomat}
\end{figure}
\subsection{Domain growth by Ostwald ripening}

 For a conserved order parameter, the dynamics of the order parameter is described by a  Cahn-Hilliard equation \cite{bray:1994}:
\be
\partial_t \phi = D\boldsymbol{\nabla}^2\mu/\kT \qquad \mu =a^2\delta \mathcal{F} /\delta \phi\ , 
\label{cahnhilliard}
\ee
where $\mu$ is the chemical potential associated with $\phi$, and $D$ is the monomer diffusion coefficient. Within the two-phases region of \fig{spinomat}, \eq{cahnhilliard} produces domains with sharp boundaries. Domains larger than a critical size $R_c$ grow at the expense of smaller domains, and inter-domain exchange occurs by monomer diffusion through the bulk phase. 

Provided diffusion is much faster than the evaporation of the smallest domains,  Lifschitz, Slyozov and Wagner (LSW) have shown that the characteristic domain size should obey the scaling law  $R_c \sim t^{1/3}$ \cite{lifshitz:1961,wagner:1961}. This can be qualitatively explained:  if diffusion is fast, the order parameter profile outside the domain boundary satisfies the quasi-static approximation $\boldsymbol{\nabla}^2\mu=0$, and the chemical potential inside domains is related to the line tension by $\mu \approx \gamma / R_c$ \cite{bray:1994}. If there is only one length scale $R_c$ in the system, mass conservation implies $\dot R_c\sim \nabla\mu\sim \mu/R_c$. Identifying ${\bf \nabla.}$ with $1/R_c$, one finds (see S.I.):
\begin{eqnarray}
R_c^2 \dot{R_c} \propto D \gamma a^2/\kT \label{rcsigma} \, ,
\label{LS}
\end{eqnarray}
leading to the classical Lifschitz-Slyozov-Wagner (LSW) dynamical scaling $R_c \sim t^{1/3}$ \cite{lifshitz:1961,wagner:1961}. The size distribution of domains growing by Ostwald ripening is strongly peaked around the size $R_c$ \cite{bray:1994}. Though this was strictly shown in dimension $3$ or higher, it has been confirmed numerically in two dimension \cite{huse:1986}.

  A similar scaling has been observed numerically {\em at steady state} in the presence of reversible reaction $A \rightleftarrows B$, time being replaced by the inverse of the reaction rate \cite{glotzer:1995}. In the following we show analytically and numerically that a similar scaling  also exists for  irreversible reactions.

\subsection{Domain growth by coalescence}

 The role of hydrodynamics on phase separation in fluid membranes is still controversial, despite considerable recent attention \cite{camley:2011,fan:2010}. For off-critical mixtures ($\phi\ne 1/2$), hydrodynamic correlations result in the diffusion and coalescence of entire domains. At the scaling level \cite{bray:1994}, the area of the largest domain can at most double at each coalescence event: $\dot R_c \le R_c/\tau_D$. The typical collision time $\tau_D=L^2/D_d$ depends on the domain diffusion coefficient  $D_d$ and the typical area per domain $L^2\sim R_c^2/\bar\phi$. Finally, one finds :
\be
R_c\dot R_c\propto D_d\bar\phi\ .
\label{hydro}
\ee
If viscous dissipation is mostly due to membrane hydrodynamics, the domain diffusion coefficient $D_d$ is only weakly dependent on domain size \cite{saffman:1975}. One thus expects $R_c\sim t^{1/2}$ for constant composition, which dominates Ostwald ripening at long times.

The size distribution of domains can be studied using the Smoluchowski coagulation equation \cite{smo:1916}:
\begin{eqnarray}
\partial_t C_n=J_n-kC_n N+\frac{k}{2}\sum_{m=1}^{n-1}C_{m}C_{n-m}\cr {\rm with} \quad N=\sum_{m=1}^\infty C_m
\label{master}
\end{eqnarray}
where $C_n$ is the concentration of domains containing $n$ monomers (if $R$ is the domain size: $n\sim R^2$), $k$ is a typical diffusion rate, and where domain scission has been neglected. This model has been studied extensively for different forms of the diffusion rate \cite{davies:1999}. Following the assumption that the diffusion coefficient of a domain is independent of its size \cite{saffman:1975}, we choose a constant diffusion rate $k=D_d/a^2$. $J_n$ is a source and sink term allowing for the creation or removal of domains \cite{turner:2005}.

In the absence of maturation ($\bar\phi=$const., $J_n=0$), the size distribution is well approximated by an exponential with a characteristic domain size $\bar n\sim \bar\phi kt$, giving the domain radius $\bar R(t)\sim \sqrt{D_d\bar\phi t}$, in agreement with \eq{hydro}.

The domain size distribution is modified by the presence of sources and sinks. It has been shown in \cite{turner:2005} that choosing a source and sink term that conserves the average concentration  $\bar\phi$ (i.e. $J_n=j_{in}\delta_{n,1}-k_{\rm off}C_n$) produces a steady-state power-law distribution $C_n\sim n^{-3/2}$, up to a characteristic domain size beyond which the distribution is exponential. The characteristic size obeys a scaling reminiscent of \eq{hydro}: $R_c\propto \sqrt{D_d\bar\phi/k_{\rm off}}$.

\section{Transient phase separation under irreversible maturation}

Maturation corresponds to the increase of $\bar{\phi}$ with time from $\bar{\phi}=0$ to $\bar{\phi}=1$ due to a chemical reaction. If $K>4 \kT$, the spinodal line (\fig{spinomat}) is crossed twice, first at $\bar{\phi}=\phi_a$ when phase separation starts, then at $\bar{\phi}=\phi_b$ where the system tends to be homogenous once again . 
We ask whether domains larger than a critical budding size, for instance given by \eq{range}, can form during the time the system is prone to phase separation ({\em i.e.} while $\phi_a <\bar\phi <\phi_b$). 

\subsection{Dynamical scaling}
 
For Ostwald ripening, the quasi-static approximation above assumes that the order parameter profile outside domains adjusts quasi-statically to domain growth ($\boldsymbol{\nabla}^2\phi=0$ in the bulk). In a maturing membrane,
the average membrane composition evolves according to $\partial_t \bar\phi= k_r (1-\bar\phi)$, and  the quasi-static composition profile satisfies :
\be
D\boldsymbol{\nabla}^2\phi+k_r(1-\phi)=0 \, ,\ee
 which defines a characteristic length scale $\lambda_D=\sqrt{D/k_r}$.  
For small domains $R\ll\lambda_D$, 
maturation does not modify the concentration profile, but merely changes the mean density outside the domains. Thus \eq{LS} should hold, with a transport coefficient now depending on time through the mean concentration $\bar\phi$. This is shown with more details in the S.I.

In the regime dominated by coalescence, \eq{hydro}, which assumes a single characteristic domain size, may be used with a time-dependent $\bar\phi$ under the assumption that the number of domains varies little between two coalescence events. This approximation is shown to be valid below.

The extent of phase separation can be characterized by the maximum size $R_{max}$ a domain can reach during the transient phase separation. Here, we are interested in membrane domains that may undergo budding, namely domains of the minority phase surrounded by the majority phase. Domains are thus of the mature species below $\phi=1/2$ and of the immature species for $\phi>1/2$ (see insets \fig{spinomat}), and the maximum domain size  occurs for $\phi=1/2$. Integrating \eq{powerlaw}, one finds:
\begin{equation}
R_{max}^{1/\alpha} \propto  \int_{\phi_a}^{1/2}  \tilde D (\bar{\phi}) \frac{d\bar\phi }{\dot{\bar\phi}}=  \frac{1}{k_r} \int_{\phi_a}^{1/2} \frac{\tilde D(\bar\phi)}{1-\bar{\phi}} d\bar{\phi} .
\label{scaling}
\end{equation}
The {\em maximum}  size of transient domains in a maturing membrane is thus predicted to follow the dynamical scaling law observed for domain growth under fixed composition (\eq{powerlaw}), where the maturation rate replaces $1/t$: $R_{max}\sim k_r^{-\alpha}$. This dynamical scaling is reminiscent of the scaling observed {\em at steady state}  in the presence of reversible reaction $A \rightleftarrows B$ \cite{glotzer:1995} or continuous recycling \cite{turner:2005}. That it is  also applicable to irreversible reactions is remarkable, since the kinetics of domain growth in a membrane undergoing maturation do not follow the same scaling, as $\bar\phi$ changes with time. This kinetics is not easily obtained from scaling arguments for Ostwald ripening, as the $\bar\phi$ dependence of \eq{LS} is not straightforward (see S.I.). For domain coalescence, a simple integration of \eq{hydro} with $\bar\phi\simeq k_r t$, valid at early time, shows that one expects $R_c(t)\simeq\sqrt{D_d k_r}t$. This linear growth contrasts with the $\sim t^{1/2}$ scaling in the absence of maturation.

 \begin{figure}[t] 
   \includegraphics[width=8.5cm]{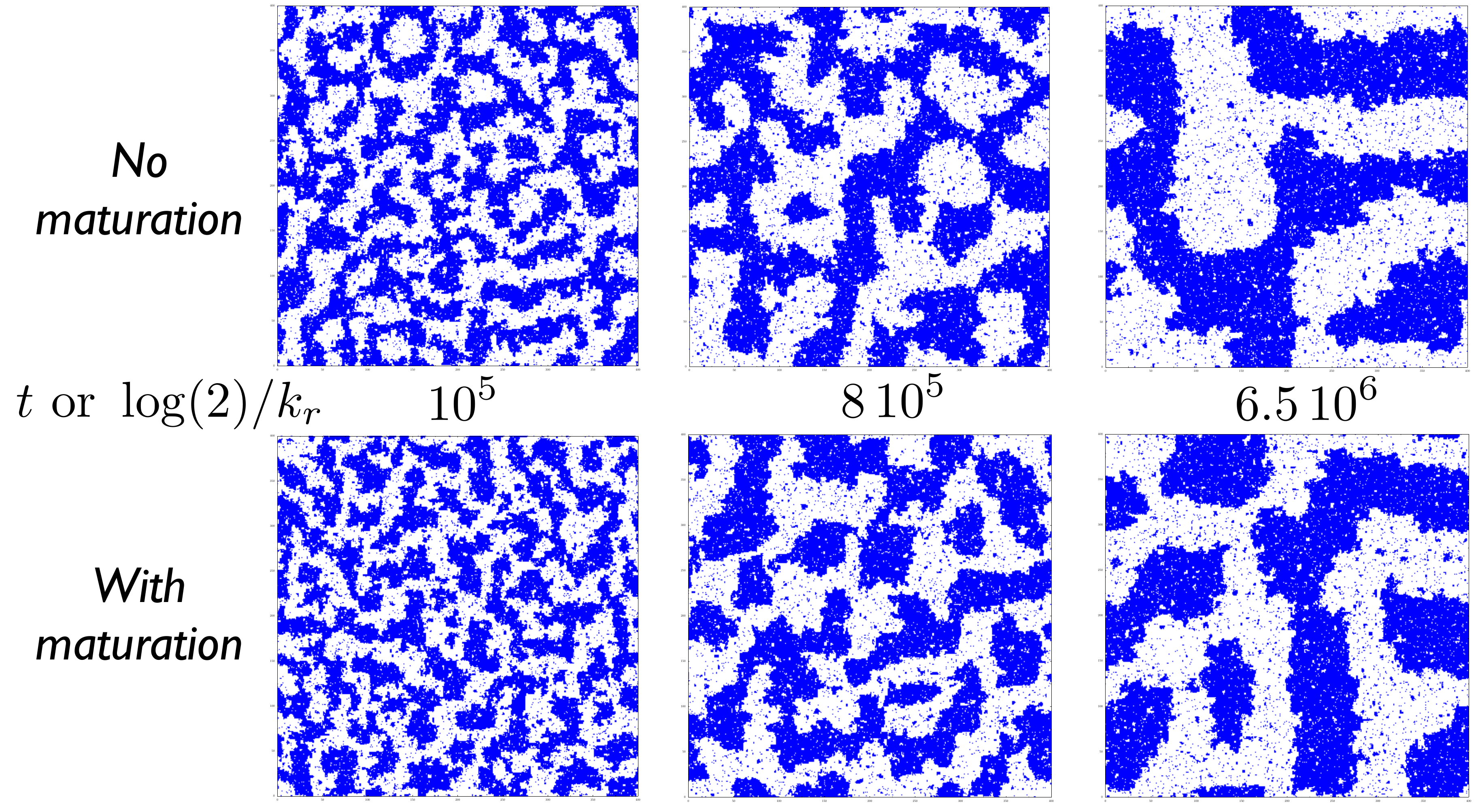} 
   \caption{\small Growth by Ostwald ripening. Snapshots of Monte Carlo simulations of domain formation in an Ising model without maturation for different times (top row) and with maturation for different maturation rates (bottom row). The average concentration is $\bar\phi=1/2$ in all cases, 
   and the interaction parameter  $J=0.75 \kT$ corresponds to a physiological line tension $\gamma\simeq4\unit{pN}$ (see S.I.)}
   \label{num}
\end{figure}

\subsection{Numerical results - Ostwald ripening}

\begin{figure}[b] 
   \includegraphics[width=8.5cm]{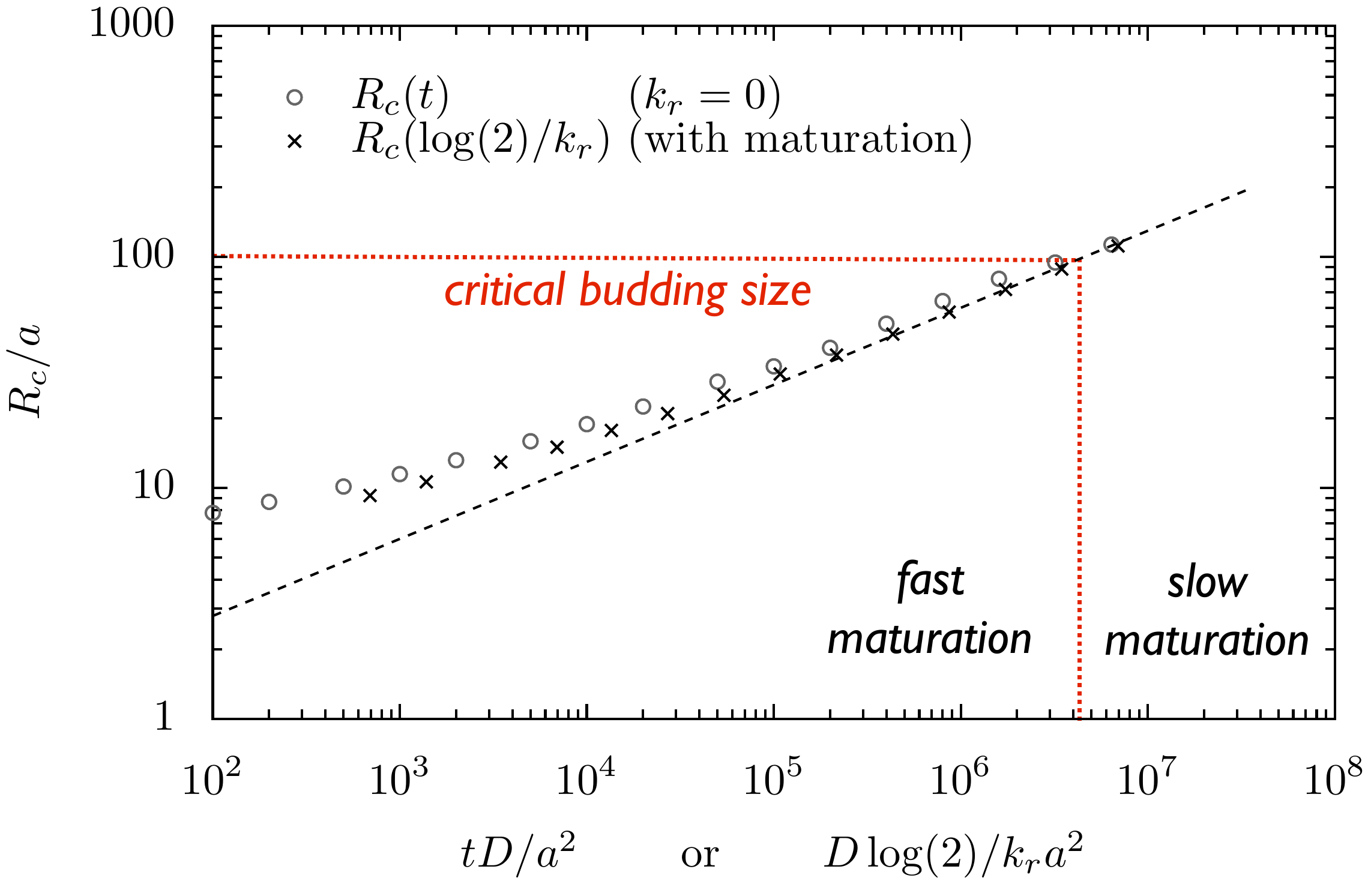} 
   \caption{\small Growth by Ostwald ripening. Domain size (in units of the molecule size $a$) as a function of time ($\circ$) and maximum domain size as function of the inverse maturation rate $1/k_r$ ($\times$), from simulations with $J=0.75$.  Time is in units of the diffusion time $a^2 / D$ ($\approx 10^{-5} s$). The dashed line corresponds to $k_r^{-1/3}$.  The critical budding size is estimated based on the line tension driven budding scenario (\eq{range})}
   \label{Lct}
\end{figure}

The prediction of \eq{scaling} was tested numerically (simulation details are given in the S.I.). Without hydrodynamics, we performed Monte Carlo simulations of the Ising model with nearest-neighbor interaction (parameter $J$) and a discrete order parameter $s$ ($0$ or $1$). Monomer diffusion is implemented using Kawasaki dynamics (spin exchange between nearer neighbors), known to produce the LSW growth in a system without maturation \cite{huse:1986,2005:krzakala}.  Maturation is implemented by letting each site with $s$=0 become a $s$=1 site with a probability $k_r dt$ at each time step (of duration $dt$). Snapshots of the simulations, shown in \fig{num}, highlight the similarity between a system without maturation (fixed $\bar\phi$) after a time $t$, and a system which reaches the same concentration by maturation (after a time $t=-\log(1-\bar\phi)/k_r$, shown for $\bar\phi=1/2$). 
\fig{Lct}  compares the variation of the average domain size with time in a system without maturation (with $\bar\phi=1/2$) and the variation of the maximum domain size with respect to the inverse maturation rate in a system undergoing maturation.
Both the LSW scaling without maturation ($R\sim t^{1/3}$), and our predicted scaling with maturation ($R_{max}\sim k_r^{-1/3}$) are apparent at late stages. Strikingly, prefactors of the power law appear very similar with or without maturation. 
 
\subsection{Numerical results - Diffusion and coalescence}

Growth by coalescence was studied numerically by solving the Master equation \eq{master} with the source term $J_n=\delta_{n,1} k_r(1-\bar\phi)$ (with $\bar\phi=\sum_n nC_n$), corresponding to monomers being constantly  ``created'' by maturation (see S.I. for details of the numerical scheme). The system contains no mature components at $t=0$ ($C_n(t=0)=0\ \forall\ n$). We find that the domain size distribution crosses over from a power-law $C_n=A n^{-3/2}$ for small $n<n^*(t)$ to an exponential decay over the size $n^*$ for large $n$. This result indicates that, in a system undergoing irreversible maturation, the domain size distribution is essentially stationary up to the cross-over size $n^*(t)$, computed analytically below. 

\begin{figure}[b] 
\includegraphics[width=3in]{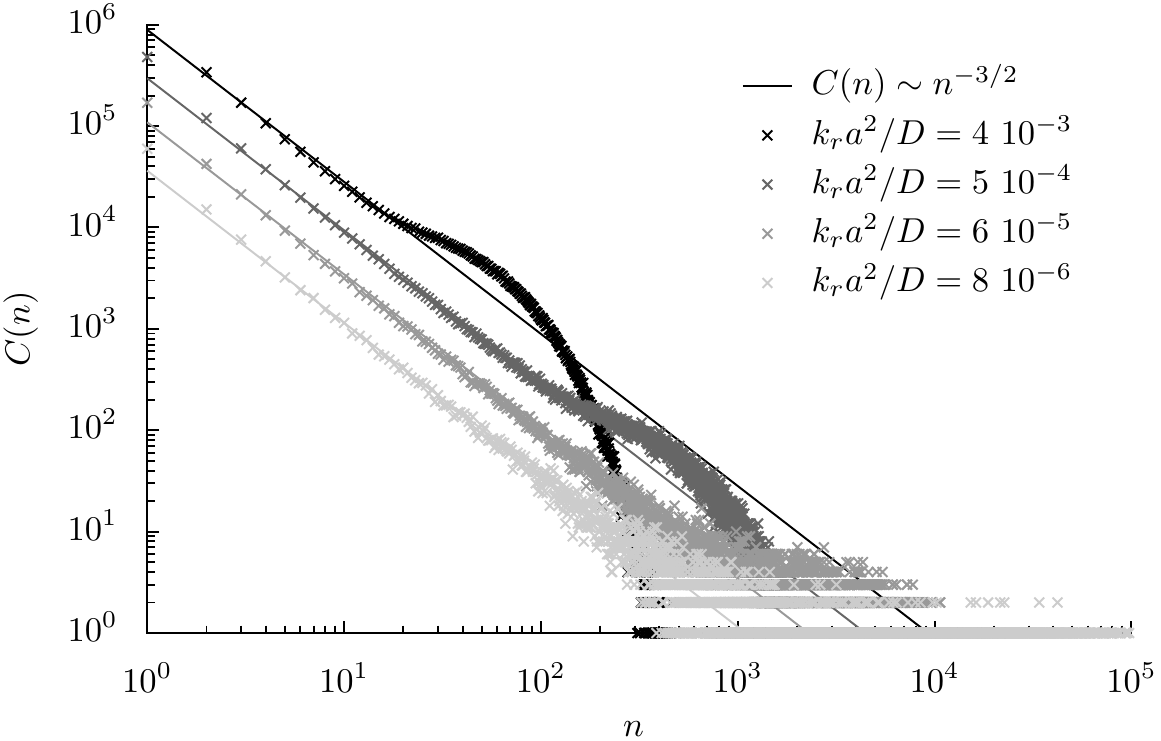}   
\caption{Growth by coalescence.	Distribution of domain size at $\bar\phi=1/2$  for different maturation rates, obtained by numerically solving  \eq{master}.  The log-log plot shows a power law behavior $\sim n^{-3/2}$ for small domains, as expected from scaling arguments. }
   \label{fig4}
\end{figure}
\begin{figure}[t] 
   \includegraphics[width=3in]{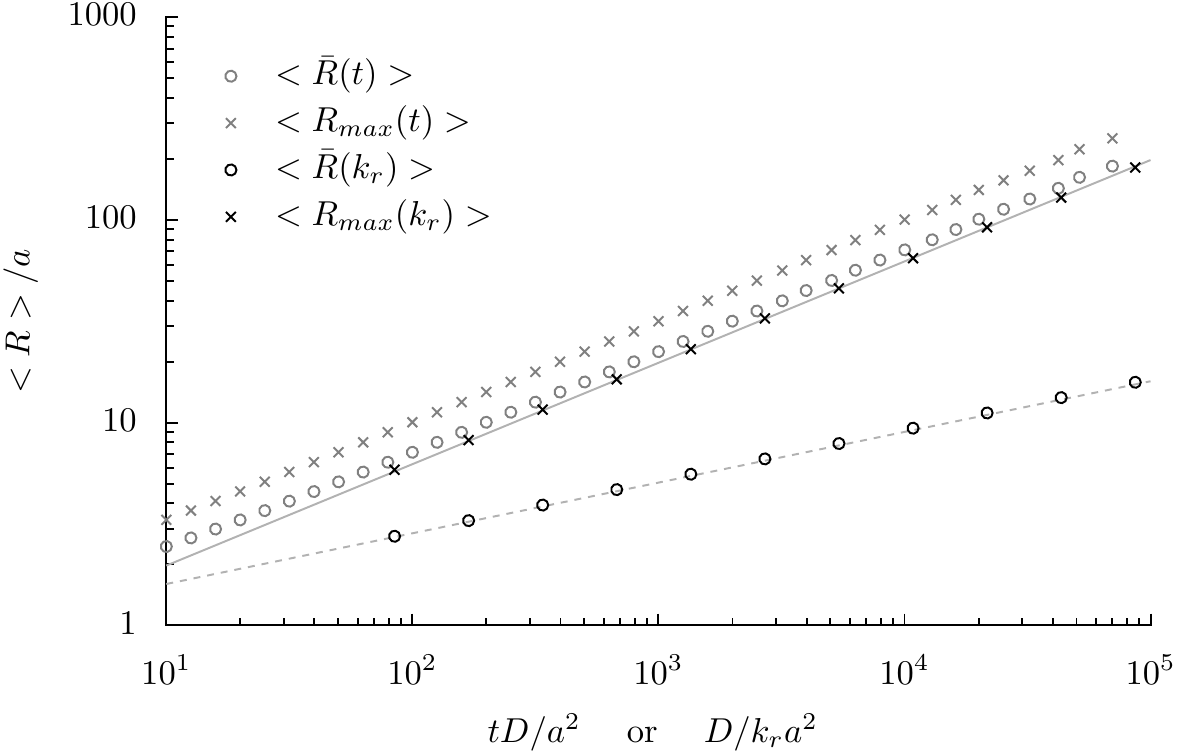}
   \caption{	Growth by coalescence. Mean domain size ($\bar{R}(t)$, grey circles)  and maximum domain size ($R_{max}(t)$, grey crosses) as a function of time in a system without maturation (constant $\bar\phi$), and mean domain size ($\bar{R}(k_r)$, black circles) and maximum domain  size ($R_{max}(k_r)$, black crosses) as a function of the inverse maturation rate $1/k_r$ in the presence of maturation. The dashed line is $\sim t^{1/4}$ and the solid line is $\sim t^{1/2}$.}
   \label{fig5}
\end{figure}

We first focus on the role of the maturation rate by analyzing domain size distribution for different values of $k_r$, when $\bar\phi$ reaches $1/2$ ($t=\log[2]/k_r$), for which phase separation is most pronounced and one expects to observe the largest domains. \fig{fig4} shows the domain size distribution for $\bar\phi=1/2$ for different values of the maturation rate. The power-law $C_n=A n^{-3/2}$ is confirmed up to a characteristic size that depends on $k_r$. This size may be computed as follows. Using \eq{master} for $n=1$, stationarity of the monomer concentration $C_1$ imposes  $A\simeq k_r(1-\bar\phi)/(k N)$, with the total number of domains $N=\sum_{n=1}^{\infty}C_n\simeq \sum_{n=1}^{n^*}C_n=\sqrt{(1-\bar\phi)k_r/k}$. Using  the conservation relation $\bar\phi=\sum_{n=1}^{\infty}n C_n\simeq\sum_{n=1}^{n^*}n C_n\simeq\sqrt{(1-\bar\phi)k_r/kn^*}$, the maximum domain size is found to be: $n^*\sim k/k_r\times\bar\phi^2/(1-\bar\phi)$, and the average domain size is $\bar n\equiv\bar\phi/N=\sqrt{n^*}$. The maximum and average domain size when $\bar\phi=1/2$  are predicted to depend on the maturation rate according to
\be
R_{max}\sim\sqrt{\frac{D_d}{k_r}}\quad{\rm and}\quad \bar R \sim\left(\frac{D_d a^2}{k_r}\right)^{1/4}
\ee
These predictions are confirmed numerically in \fig{fig5} which shows the variation of these two characteristic length scales with the maturation rate.

\begin{figure}[b] 
\includegraphics[width=3in]{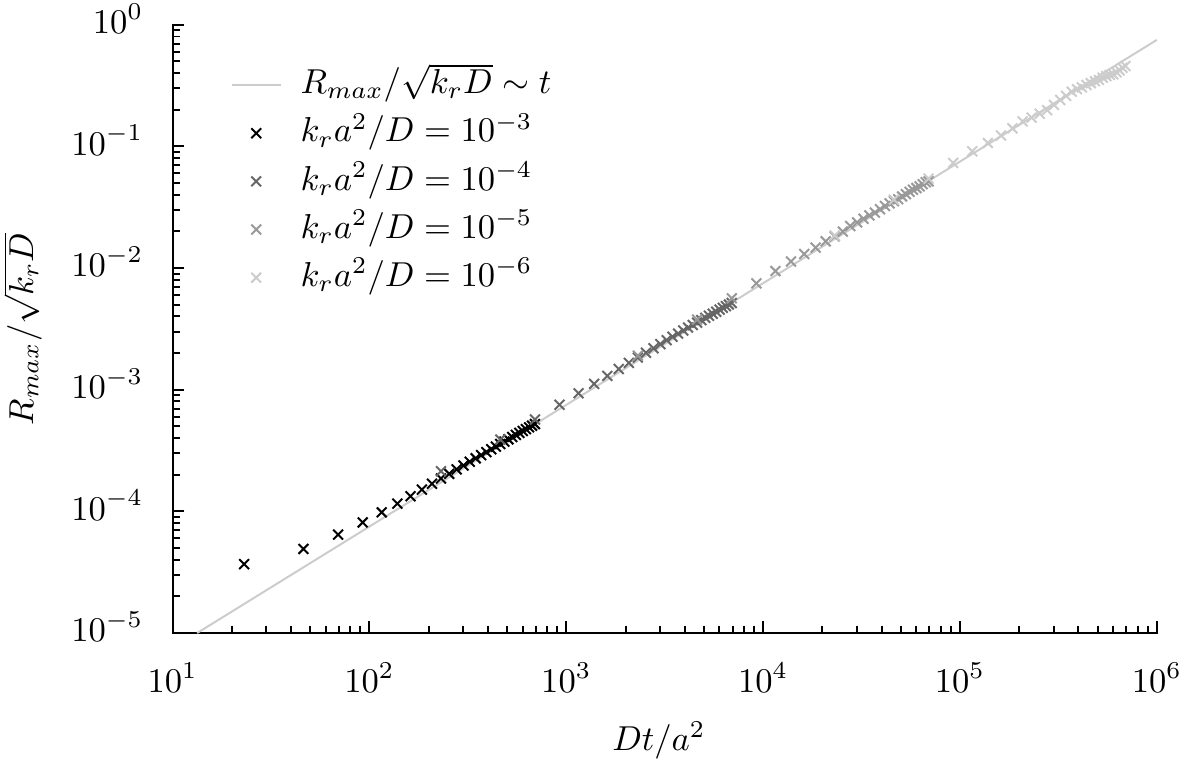}
   \caption{Growth by coalescence. Variation of the maximum domain size as a function of time when maturation is present. The linear growth law $R_{max} / \sqrt{k_r D} \sim t$ predicted in the text is observed. The maximum domain size is defined as $R_{max}=\sqrt{\sum n^2 C_n/\bar\phi}$.}
   \label{fig6}
\end{figure}

As discussed above, the dynamical scaling for domain growth can be obtained for the entire maturation process (at least while the matured species is the minority, $\bar{\phi} \in [0,1/2]$) using  \eq{hydro}: $R_c\propto\sqrt{D_d k_r}t$. The characteristic domain size is predicted to increase linearly with time due to the combined effect of domain coalescence and the increasing fraction of matured species, both accounting for $\sqrt{t}$. This prediction was  verified numerically, as shown \fig{fig6}.

\section{Discussion}

The dynamical scaling predicted by \eq{scaling} is thus universally observed  whether domain growth proceeds  by Ostwald ripening or by domain coalescence.  On time scales consistent with biochemical maturation: $1/k_r\sim\unit{min.}$, the maximum size of transient domains in a membrane undergoing irreversible maturation is $\sim0.1\unit{\mu m}$ with Ostwald ripening (\eq{LS} and \fig{Lct}) and $\sim 1 \unit{\mu m}$ by domain coalescence (\eq{hydro} and S.I.), with $D=0.1\unit{\mu m^2/s}$, $a=1\unit{nm}$, and $\gamma a/\kT\sim 1$.   
 
 On deformable membranes, domains within the size range of \eq{range}  undergo line tension-driven budding. Membrane deformability does not modify the early stages of domain growth, but has a complex influence on late-stage growth. Numerical studies report the possible fusion of budded domains \cite{sunil:1998,sunil:2001,laradji:2004}, but membrane-mediated repulsion between non-flat  domains may prevent their coalescence \cite{yanagisawa:2007,ursell:2009}. In low membrane tension organelles, domains large enough to deform will form a complete bud \cite{pio_regul} that may undergo scission, and the late-stage dynamics should be less relevant.
 
Within our framework, one may expect irreversible maturation of membrane components to lead to domain budding and irreversible morphological changes if transient domains can reach the critical budding size, a possibility that requires slow maturation rates. This prediction could be tested experimentally on artificial membrane systems (giant unilamellar vesicles). We predict that a large ($\gtrsim\unit{\mu m}$) deformable vesicle undergoing chemical maturation would preserve its integrity if the reaction is fast, while it would split in two or more daughter vesicles if the reaction is slower ($\gtrsim\unit{min.}$) and the maximum transient domain size exceeds the budding size. One possibility would be to use the sphingomyelinase-induced maturation of ceramid into sphingomyelin in giant vesicles, as this reaction is of physiological interest since it occurs in the Golgi apparatus, and is known to produce lipid domains \cite{fanani:2009}.

Extending our results to multicomponent cellular membranes is not  straightforward, since many factors may influence domain growth and budding and  participate in domain size regulation. Specific membrane proteins  promote curvature and fission \cite{mcmahon:2005} and may modify the critical budding size range compared to \eq{range}. Interaction with the cytoskeleton may prevent domain diffusion and coalescence \cite{fisher:2011,ehrig:2011}. However, transient sub-micron domains have been seen on yeast Golgi cisternae \cite{matsuura:2006,losev:2006} and slightly larger domains in mammals \cite{patterson:2008}. This suggests that domain formation is an important component impacting the dynamics of membrane-bound organelles. We thus venture the proposal that the  rate of maturation of membrane components could fundamentally affect the  morphology and dynamics of cellular organelles. 


Our study appears particularly interesting in the case of the Golgi apparatus. We argue that the two extreme Golgi organizations observed in nature can  be fitted within a single framework. Yeast Golgi (fast maturation, $k_r\sim 1/\unit{min}$) could be made of dispersed cisternae undergoing independent maturation, because the maturation rate is too fast for the emergence of membrane domains that can reach the budding size. On the other hand, the fact that the Golgi of mammalian cells is a stack of interacting cisternae of different biochemical identities ({\em cis,\ medial,\ trans}) could be made possible by a relatively slow maturation rate ($k_r\sim 1/20\unit{min}$) allowing the formation of large mature domains. Although this simple picture is far from capturing the full complexity of the Golgi apparatus, and in particular the compositional complexity present in other models \cite{heinrich:2005, gong:2008} or the recycling of resident Golgi enzymes by specific retrograde transport, our results suggest that an internal property of an organelle (the rate of chemical reaction in the Golgi apparatus) could control the structure and organization of this organelle.

\begin{acknowledgments} 
We gratefully acknowledge B. Goud, N. Gov, R. Phillips and M. Rao for stimulating discussion.  
\end{acknowledgments}

\bibliographystyle{apsrev4-1}


%

\end{document}